\documentclass[epj]{svjour}
\usepackage{graphics}
\begin{document}
\title{ Shape transformation transitions of a tethered surface model}

\author{Hiroshi Koibuchi
}                     
%
%
\institute{Department of Mechanical and Systems Engineering \\
  Ibaraki National College of Technology \\
  Nakane 866, Hitachinaka, Ibaraki 312-8508, Japan }
%
%
\abstract{
A surface model of Nambu and Goto is studied statistical mechanically by using the canonical Monte Carlo simulation technique on a spherical meshwork. The model is defined by the area energy term and a one-dimensional bending energy term in the Hamiltonian. We find that the model has a large variety of phases; the spherical phase, the planar phase, the long linear phase, the short linear phase, the wormlike phase, and the collapsed phase. Almost all two neighboring phases are separated by discontinuous transitions. It is also remarkable that no surface fluctuation can be seen in the surfaces both in the spherical phase and in the planar phase. 
}
\authorrunning {H.Koibuchi}
\titlerunning {Shape transformation transitions of a tethered surface model}
\maketitle
\section{Introduction}\label{intro}
Membranes have long been interested in biological physics and in softmatter physics \cite{NELSON-SMMS2004,Gompper-Schick-PTC-1994,Bowick-PREP2001}. The shape of membranes has been studied experimentally and theoretically \cite{Yoshikawa,Hotani,SBR-PRA1991,EVANS-BPJ1974,JSWW-PRW1995}.
The statistical mechanical studies for the conformation and the elastic property of membranes were reviewed in \cite{GK-SMMS2004}.

The curvature surface model of Helfrich and Polyakov \cite{HELFRICH-1973,POLYAKOV-NPB1986,KLEINERT-PLB1986} plays a role for describing the shape of membranes from the two-dimensional differential geometrical or the classical mechanical view point. In fact, the surface shapes are understood as the conditions that minimize the curvature energies of the models such as the area difference bilayer model and the minimal model \cite{SBR-PRA1991,EVANS-BPJ1974,JSWW-PRW1995}. It is also interesting to study how thermal fluctuations influence and change the membrane shape \cite{DS-EPL1996,PM-EPL2002}. The prolate-oblate transition was in fact observed as a phenomenon driven by thermal fluctuations \cite{DS-EPL1996}. Therefore, it is still worthwhile to study numerically the shape transformation phenomena as phase transitions in the surface models. 

Contrary to the shape transformation, the surface fluctuation phenomena were extensively studied theoretically \cite{Peliti-Leibler-PRL1985,David-Guitter-EPL1988,PKN-PRL1988} and numerically \cite{KANTOR-NELSON-PRA1987,Baum-Ho-PRA1990,CATTERALL-NPBSUP1991,AMBJORN-NPB1993,KOIB-EPJB-2005,GREST-JPIF1991,BOWICK-TRAVESSET-EPJE2001,BCTT-PRL2001} in the surface model of Helfrich and Polyakov. It was reported recently that the transition is of first-order on the fixed connectivity spherical surfaces \cite{KD-PRE2002,KOIB-PRE-2005,KOIB-NPB-2006}. We see that the transition is accompanied by the collapsing transition, which can be considered as a shape transformation transition between the smooth phase and the collapsed phase. 

Here we comment on the terminology used in this paper for membranes models. The Gaussian bond potential is defined by the summation over the bond length squares on triangulated surfaces, while the Nambu-Goto energy is defined by the summation over the area of triangles. Both potentials play a role of the microscopic surface tension in the surface models. A surface model defined by the Gaussian bond potential with an extrinsic curvature energy term is called the Helfrich and Polyakov model, while the one defined by the Nambu-Goto energy is called the Nambu and Goto model. On the contrary, the fluid surface model is the one defined on dynamically triangulated surfaces, and hence the vertices diffuse freely on such fluid surfaces. The fixed-connectivity (or tethered) surface model is defined on fixed connectivity triangulated surfaces, and hence the vertices can only fluctuate locally on such fixed-connectivity surfaces.

We must recall that the surface models for describing the surface fluctuation phenomena can be classified by the bending energy (extrinsic curvature or intrinsic curvature), the surface tension energy (Gaussian bond potential or Nambu-Goto energy), and the fluidity (fixed connectivity or fluid). Surface models show a variety of shapes, which strongly depends on these physical quantities \cite{KOIB-PRE2004,KOIB-EPJB2004}. The surface model of Nambu-Goto with an intrinsic curvature energy has a variety of phases even on fixed connectivity spherical surfaces \cite{KOIB-PRE2004}.

One additional element that characterizes the surface model is the cytoskeletal structure \cite{HHBRM-PRL-2001}, where the bending energy is not always a two-dimensional one \cite{KOIB-JSTP-2007,KOIB-EPJB-2007}. The cytoskeletal structure is a lattice analog of cytoskeletons or microtubules in biological membranes. The one-dimensional bending energy, which is defined on the skeletons of the surface, can also give the mechanical strength to the surface for maintaining the shape. We can call a surface model with such cytoskeletal structure as the compartmentalized model. It was shown that such compartmentalized models on fluid surfaces have a variety of phases including the planar phase and the tubular phase \cite{KOIB-PRE2007-2}.

Therefore, it is very interesting to study a surface model defined by combining these elements such as the Nambu-Goto energy and the cytoskeletal structure. In this paper, we study the Nambu-Goto surface model with a one-dimensional bending energy on fixed connectivity spherical surfaces. Although the lattice is a triangulated one and is identical to that for the standard model without the compartmentalized structure, the lattice is considered to be the compartmentalized one. In fact, the lattice is obtained from the finite-size compartmentalized one in the limit of $L\to 1$ (in the unit of bond length), where $L$ is the length of the compartment boundary between two junctions. The compartment size can also be characterized by $n$ the total number of vertices in a compartment. The one-dimensional bending energy is defined on the compartments of the model of finite $n$ \cite{KOIB-JSTP-2007}. In Ref.\cite{KOIB-PLA-2007}, we studied a compartmentalized model with the one-dimensional bending energy in the limit of $n\to 0$ and reported certain differences between the results of the compartmentalized model at $n\to 0$ and those of the conventional model in \cite{KOIB-PRE-2005}. It is remarkable that the results in this paper are significantly different from those in \cite{KOIB-PLA-2007}. Only difference between the model in this paper and that in \cite{KOIB-PLA-2007} is in the bond potential; the Nambu-Goto energy is assumed in the model of this paper while the Gaussian bond potential is assumed in \cite{KOIB-PLA-2007}. 

\section{Model and Monte Carlo technique}
Triangulated lattices are obtained by splitting the icosahedron, which has 12 vertices of coordination number $q\!=\!5$. By dividing every edge of the icosahedron into $\ell$ pieces, we have a triangulated lattice of size $N\!=\!10\ell^2\!+\!2$, where the generated $N\!-\!12$ vertices are of coordination number $q\!=\!6$.  

The surface model in this paper is defined by the Hamiltonian $S$, which is given by a linear combination of the Nambu-Goto energy $S_1$ and the one-dimensional bending energy $S_2$ such that
\begin{equation}
\label{S1S2}
S=S_1+b S_2, \quad S_1=\sum_{\it \Delta} A_{\it \Delta},\quad S_2=\sum_{ij} (1-{\bf t}_i \cdot {\bf t}_j),
\end{equation}
where $b$ is the bending rigidity. $A_{\it \Delta}$ in $S_1$ is the area of the triangle ${\it \Delta}$, and ${\bf t}_i$ in $S_2$ is a unit tangential vector of the bond $i$. $\sum_{\it \Delta}$ in $S_1$ is the sum over all triangles ${\it \Delta}$, and $\sum_{ij}$ in $S_2$ is the sum over bonds $i$ and $j$. 

\begin{figure}[hbt]
\unitlength 0.1in
\begin{picture}( 0,0)(  20,28)
\put(21,28.5){\makebox(0,0){(a)} }%
\put(38,28.5){\makebox(0,0){(b)} }%
\end{picture}%
\centering
\resizebox{0.49\textwidth}{!}{%
\includegraphics{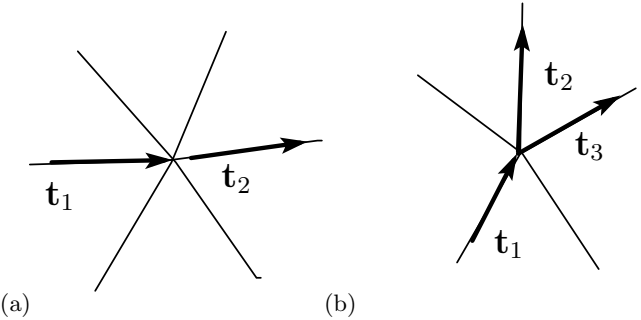}
}
\caption{(a) A combination $1\!-\!{\bf t}_1 \cdot {\bf t}_2$ is included in the bending energy with the weight of $1$ at the $q\!=\!6$ vertices,  and (b) a combination $1\!-\!{\bf t}_1 \cdot ({\bf t}_2\!+\!{\bf t}_3)/2$ is included in the bending energy with the weight of $1/2$ at the $q\!=\!5$ vertices. }
\label{fig-1}
\end{figure}

In $S_2$, the combination $1\!-\!{\bf t}_i \cdot {\bf t}_j$ of the vectors ${\bf t}_i$ and ${\bf t}_j$ is defined as follows: Three combinations are defined at the $q\!=\!6$ vertices, and $5/2$ combinations are defined at the $q\!=\!5$ vertices. Figures \ref{fig-1}(a) and \ref{fig-1}(b) show the combinations $1\!-\!{\bf t}_1 \cdot {\bf t}_2$ and $1\!-\!{\bf t}_1 \cdot ({\bf t}_2\!+\!{\bf t}_3)/2$, which are respectively typical of the $q\!=\!6$ vertices and of the $q\!=\!5$ vertices. The combination $1\!-\!{\bf t}_1 \cdot {\bf t}_2$ is included in the summation of $S_2$ with the weight of $1$, while the combination $1\!-\!{\bf t}_1 \cdot ({\bf t}_2\!+\!{\bf t}_3)/2$ is included in $S_2$ with the weight of $1/2$. This is the reason why the $q\!=\!5$ vertices have $5/2$ combinations. 

The partition function of the model is defined by 
\begin{equation} 
\label{Part-Func}
 Z =\int^\prime \prod _{i=1}^{N} d X_i \exp\left[-S(X)\right],
\end{equation} 
where $X_i (\in {\bf R}^3)$ is the position of the vertex $i$ on the triangulated spherical surface.

We should note on the problem of well-definedness of the Nambu-Goto surface model. Our model in this paper will be included into a class of well-defined models of Nambu-Goto. The surface model with Hamiltonian $S\!=\!S_1$ is known to be ill-defined \cite{ADF-NPB1985}. In fact, it is easy to see numerically that the surface is just like a chestnut bur in the model defined by $S\!=\!S_1$. Since the spines are extremely long, the resulting surface appears to be a gas. Some of the vertices are spread over large space in ${\bf R}^3$. This ill-definedness persists even when the Hamiltonian includes the standard bending energy of the form $1\!-\!{\bf n}_i\cdot {\bf n}_j$, where ${\bf n}_i$ is a unit normal vector of the triangle $i$. On the contrary, the model turns to be a well-defined one if certain types of bending energies are included in the Hamiltonian \cite{KOIB-NPB-2006,KOIB-PRE2004}. A bending energy, which is an extrinsic curvature, makes the model well-defined, and it was shown that the model has the smooth phase and the collapsed phase \cite{KOIB-NPB-2006}. It was also reported that an intrinsic curvature energy makes the model well-defined, and a variety of phases can be seen in the model \cite{KOIB-PRE2004}. 

The three-dimensional integrations in the partition function are simulated by the random three-dimensional shift of $X$ in the canonical Monte Carlo (MC) simulations. The new position $X^\prime$ is given by $X^\prime\!=\!X\!+\!\delta X$, where $\delta X$ is randomly chosen in a small sphere. The radius of the small sphere is fixed at the beginning of the simulation to maintain about $50\%$ acceptance rate.

The surface size assumed in the simulations is $N\!=\!1442$,  $N\!=\!2562$, and  $N\!=\!4842$. The total number of Monte Carlo sweeps (MCS) after the thermalization MCS is about $1\!\times\!10^8\sim 2\!\times\!10^8$ for the surfaces of size $N\!=\! 1442$,  $2\!\times\!10^8\sim 3\!\times\!10^8$ for the surfaces of size $N\!=\! 2562$, $N\!=\! 4842$ at $b$ close to the transition point, and relatively smaller number of MCS is assumed at $b$ far from the transition point.

\section{Results}
\begin{figure}[hbt]
\unitlength 0.1in
\begin{picture}( 0,0)(  00,00)
\put(1,37.5){\makebox(0,0){(a)} }%
\put(17,37.5){\makebox(0,0){(b)} }%
\put(1,-0.8){\makebox(0,0){(c)} }%
\put(17,-0.8){\makebox(0,0){(d)} }%
\end{picture}%
\centering
\resizebox{0.46\textwidth}{!}{%
\includegraphics{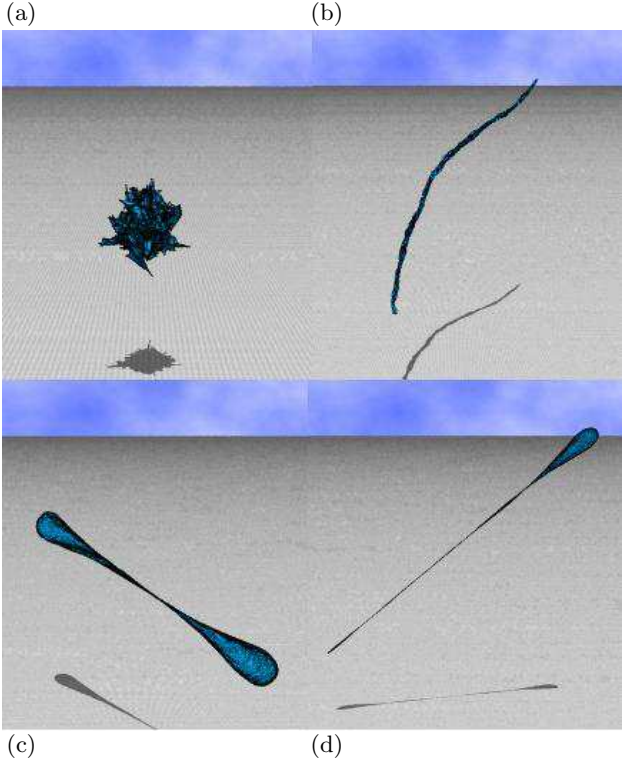}
}
\vspace{0.3cm}
\caption{Snapshots of surfaces obtained at (a) $b\!=\!0.5$ (collapsed phase), (b) $b\!=\!1$ (wormlike phase), (c) $b\!=\!30$ (short linear phase), and (d) $b\!=\!120$ (long linear phase).  The surface size is $N\!=\!4842$. The scale of the figures is different from to each other. }
\label{fig-2}
\end{figure}
\begin{figure}[hbt]
\unitlength 0.1in
\begin{picture}( 0,0)(  0,0)
\put(1,37.5){\makebox(0,0){(a)} }%
\put(17,37.5){\makebox(0,0){(b)} }%
\put(1,-0.9){\makebox(0,0){(c)} }%
\put(17,-0.9){\makebox(0,0){(d)} }%
\end{picture}%
\centering
\resizebox{0.46\textwidth}{!}{%
\includegraphics{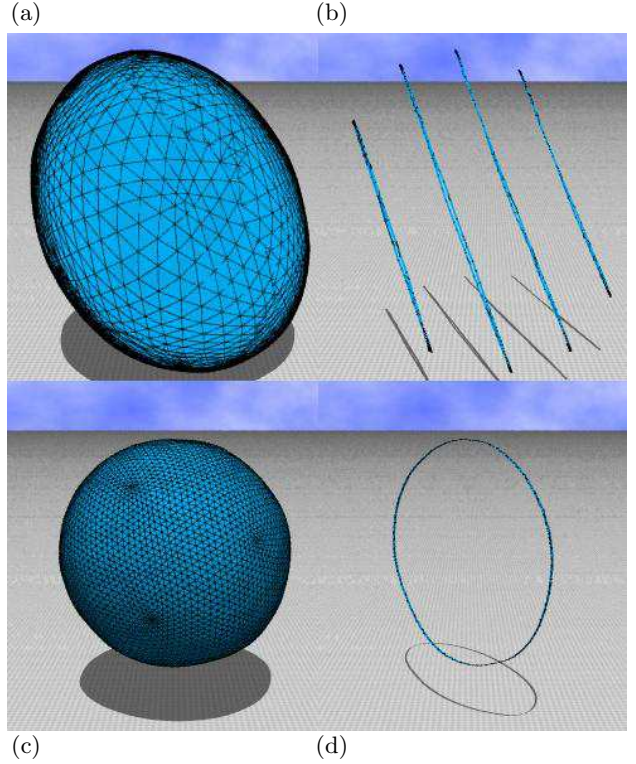}
}
\vspace{0.3cm}
\caption{Snapshots of surfaces and their surface sections of size $N\!=\!4842$ obtained  at (a),(b) $b\!=\!140$ (planar phase) and at (c),(d) $b\!=\!190$ (spherical phase). These figures were drawn in the same scale, which is also identical to that of Fig.\ref{fig-2}(a). }
\label{fig-3}
\end{figure}
First we show snapshots of surfaces of size $N\!=\!4842$ in Figs.\ref{fig-2}(a)--\ref{fig-2}(d) respectively obtained at $b\!=\!0.5$ (collapsed phase), $b\!=\!1$ (wormlike phase), $b\!=\!30$ (short linear phase), and $b\!=\!120$ (long linear phase). The scales are different to each other figure, because the size of surfaces extends from a small scale to a very large one as we will see in the mean square size later in this section. We find from Figs.\ref{fig-2}(c) and \ref{fig-2}(d) that the surfaces in the short and the long linear phases are almost one-dimensional except the two terminal points, which have a two-dimensional structure. 

Figures \ref{fig-3}(a) and \ref{fig-3}(b) are the snapshot and the surface section obtained at $b\!=\!140$ (planar phase), and Figs \ref{fig-3}(c) and \ref{fig-3}(d) are those obtained at $b\!=\!190$ (spherical phase). The snapshots were drawn in the same scale, which is also identical to that of Fig.\ref{fig-2}(a).
From the section in Fig.\ref{fig-3}(b) of the planar surface, we understand that the surface is flat in the planar phase. It is also seen that the surfaces are quite smooth in both phases. The smoothness is in sharp contrast to the case of the standard surface model such as the one in \cite{KOIB-PRE-2005}, where the surface is not so smooth even in the smooth phase.

\begin{figure*}[hbt]
\centering
\resizebox{0.7\textwidth}{!}{%
\includegraphics{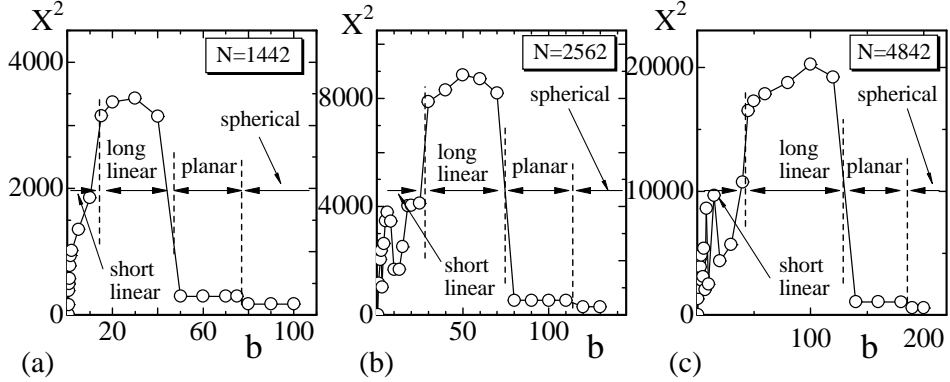}
}
\caption{The mean square size $X^2$ versus $b$ obtained on the surfaces of size (a) $N\!=\!1442$, (b) $N\!=\!2562$, and (c) $N\!=\!4842$. Dashed lines denote the phase boundaries. }
\label{fig-4}
\end{figure*}
The mean square size $X^2$ is defined by 
\begin{equation}
\label{X2}
X^2={1\over N} \sum_i \left(X_i-\bar X\right)^2, \quad \bar X={1\over N} \sum_i X_i,
\end{equation}
where $\sum_i$ denotes the summation over the vertices $i$, and $\bar X$ is the center of the surface. The size and the shape of surfaces can be reflected in $X^2$.

Figures \ref{fig-4}(a), \ref{fig-4}(b), and \ref{fig-4}(c) show $X^2$ vs. $b$ of the $N\!=\!1442$ surface, the $N\!=\!2562$ surface, and the $N\!=\!4842$ surface, respectively. Dashed lines drawn vertically in the figures denote the phase boundaries, where $X^2$ discontinuously changes. The discontinuous changes of $X^2$ indicate that 
the transitions are of first-order. The phase boundary between the collapsed phase and the wormlike phase and that between the wormlike phase and the short linear phase are located in the small $b$ region and therefore, they are not depicted in Fig.\ref{fig-4}. We can also see that $X^2$ wildly varies in the short linear phase on the surfaces of $N\!=\!2562$ and $N\!=\!4842$ in Figs.\ref{fig-4}(b) and \ref{fig-4}(c). $X^2$ is sensitive to the size of the two-dimensional parts at the two terminal points of the surface in the short linear phase, and the size varies in that phase. For this reason, $X^2$ wildly varies in the short linear phase.

\begin{figure*}[hbt]
\centering
\resizebox{0.69\textwidth}{!}{%
\includegraphics{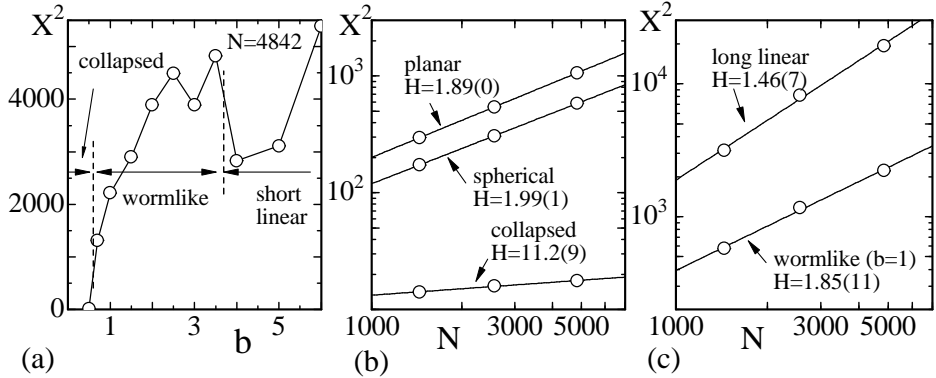}
}
\caption{(a) The mean square size $X^2$ versus $b$ in the small $b$ region, (b) log-log plots of $X^2$ against $N$ at the collapsed phase, at the planar phase and at the spherical phase, and (c) log-log plots of $X^2$ against $N$ at the wormlike phase and the long linear phase. }
\label{fig-5}
\end{figure*}
In order to see the phase structure in the small $b$ region, we plot in Fig.\ref{fig-5}(a) $X^2$ versus $b$ obtained at this region. The dashed lines were drawn at the phase boundary between the collapsed phase and the wormlike phase and at the boundary between the wormlike phase and the short linear phase of the $N\!=\!4842$ surface. We find that $X^2$ continuously varies at the boundary between the collapsed phase and the wormlike phase. This indicates that the transition between the two phases is not of first-order but of higher-order. On the contrary, the wormlike phase and the short linear phase are connected by a first-order transition, because we see that $X^2$ discontinuously changes at the boundary.

Figure \ref{fig-5}(b) shows log-log plots of $X^2$ against $N$ obtained in the collapsed phase at $b\!=\!0.5$, in the planar phase at $b$ close to the spherical phase, and in the wormlike phase at $b$ close to the planar phase. Moreover, we show in Fig.\ref{fig-5}(c)  log-log plots of $X^2$ against $N$ obtained in the long linear phase at $b$ close to the planar phase, and in the wormlike phase at $b\!=\!1$. 

The straight lines were drawn by fitting the data to
\begin{equation}
\label{Hausdorff}
X^2 \sim N^{2/H},
\end{equation}
where $H$ in the exponent is called the Hausdorff dimension; the smoothness and the shape of surface can be reflected in $H$. Thus, we have $H_{\rm col}$ (collapsed phase), $H_{\rm wor}$ (wormlike phase), $H_{\rm lli}$ (long linear phase), 
$H_{\rm pla}$ (planar phase), and $H_{\rm sph}$ (spherical phase), which are shown in Table \ref{table-1}.
\begin{table}[hbt]
\caption{The Hausdorff dimension $H$ obtained in the collapsed phase, in the wormlike phase, in the long linear phase, in the planar phase, and in the spherical phase.}
\label{table-1}
\begin{center}
 \begin{tabular}{cccccc}
\hline
$H_{\rm col}$ & $11.2\pm0.89$  &&& $H_{\rm wor}$ & $1.85\pm0.11$  \\
 $H_{\rm lli}$ & $1.46\pm0.07$  &&& $H_{\rm pla}$ & $1.89\pm0.00$ \\
 $H_{\rm sph}$ &  $1.99\pm0.01$ &&&& \\
 \hline
 \end{tabular} 
\end{center}
\end{table}

We note that $H_{\rm sph}\!=\!1.99(1)$ is almost identical to the topological dimension of surfaces, and this result is consistent to the fact that the surface in the spherical phase is smooth and spherical as shown in Fig.\ref{fig-3}(c). On the other hand, we see that $H_{\rm pla}\!=\!1.89(0)$ slightly deviates from the expected value $H\!=\!2$, which is obtained under the uniform distribution of vertices on the disk. The reason of the deviation of $H_{\rm pla}$ from $H\!=\!2$ seems that the distribution of vertices is non-uniform as can be seen in the snapshot in Fig.\ref{fig-3}(a), where vertices are dense in the perimeter region while it is sparse in the central part of the surface.  
 
$H_{\rm lli}\!=\!1.46(7)$ also deviates from $H\!=\!1$, which is the Hausdorff dimension of the straight line of uniform density of vertices. The surface in the long linear phase is not exactly identical to the straight line, because the surface includes the two-dimensional parts at the two terminal points. The surfaces in the wormlike phase are not exactly one-dimensional, and therefore the Hausdorff dimension is expected to deviate from $H\!=\!1$. This is a reason why the results $H_{\rm wor}\!=\!1.85(11)$ deviates from the expectation.

Finally, we comment on the Hausdorff dimension in the collapsed phase. The reason of $H_{\rm col}\!=\!11.2(89)$ is larger than the physical bound $H\!=\!3$ is that the surface is collapsed in the collapsed phase; the surface is allowed to self-intersect and is a phantom surface. If $X^2$ remained constant when $N$ increased, $H$ can be $H\to \infty$. However, in the model of this paper $X^2$ gradually increases with increasing $N$ as we see in Fig.\ref{fig-5}(b). This small and steady increase of $X^2$ against $N$ indicates that $H$ has a finite value. We note also that the result $H_{\rm col}\!>\!3$ is not always typical of the Nambu-Goto surface model. In fact, the Hausdorff dimension in the collapsed phase is larger than the physical bound; $H\!>\!3$, in surface models with an intrinsic curvature energy \cite{KOIB-EPJB2004}, where the surface tension energy is the Gaussian bond potential.

\begin{figure*}[hbt]
\centering
\resizebox{0.7\textwidth}{!}{%
\includegraphics{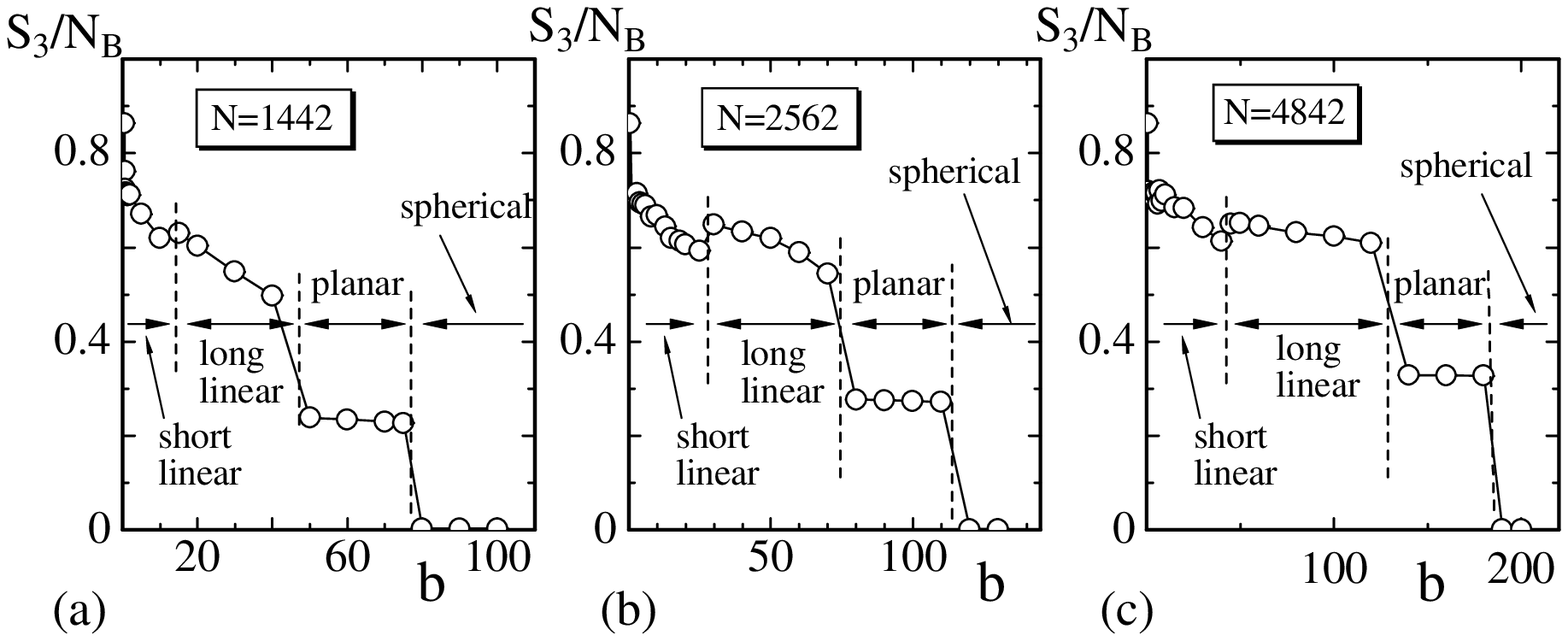}
}
\caption{The two-dimensional bending energy $S_3/N_B$ versus $b$ obtained on the surfaces of size (a) $N\!=\!1442$, (b) $N\!=\!2562$, and (c) $N\!=\!4842$. Dashed lines denote the phase boundaries.  }
\label{fig-6}
\end{figure*}
The two-dimensional bending energy $S_3$ is defined by
\begin{equation}
\label{S3}
S_3=\sum_{ij} (1-{\bf n}_i \cdot {\bf n}_j),
\end{equation}
where ${\bf n}_i$ is a unit normal vector of the triangle $i$. We should note that $S_3$ reflects the bending deformation, and therefore $S_3$ changes its value depending on how wildly the surface fluctuates, although $S_3$ is not included in the Hamiltonian. Therefore, the phase transition of shape transformation can also be characterized by anomalous behaviors of $S_3$, i.e., the discontinuous change of $S_3$, because the shape transformation phenomena is expected to be accompanied by the surface fluctuation phenomena as in the standard two-dimensional curvature model with Gaussian bond potential.

Figure \ref{fig-6}(a)--\ref{fig-6}(c) show the two-dimensional bending energy $S_3/N_B$ versus $b$ obtained on the surface of size $N\!=\!1442$, $N\!=\!2562$, and $N\!=\!4842$. $N_B$ is the total number of bonds, which is given by $N_B\!=\!3N\!-\!6$. The dashed lines were drawn in the figure at the same positions as those in Figs.\ref{fig-4}(a)--\ref{fig-4}(c) and therefore denote the phase boundaries. We find discontinuities of $S_3/N_B$ at the phase boundaries in Figs.\ref{fig-6}(a), \ref{fig-6}(b), and \ref{fig-6}(c). This observation also implies that the first-order transitions of surface fluctuation occur at the phase boundaries indicated by discontinuous changes of $X^2$ in Figs.\ref{fig-4}(a)--\ref{fig-4}(c). It is easy to understand that $S_3/N_B\!\simeq\!0$ in the spherical phase, because the surface is sufficiently smooth as shown in the snapshot in Fig.\ref{fig-3}(c). On the contrary, $1-{\bf n}_i \cdot {\bf n}_j$ has its maximal value at the edge of the disk in the planar phase, and $1-{\bf n}_i \cdot {\bf n}_j$ is expected be $0$ on the surface except at the edge. Thus, we understand the reason of the gap in $S_3/N_B$ at the boundary between the planar phase and the spherical phase.

\begin{figure*}[hbt]
\centering
\resizebox{0.7\textwidth}{!}{%
\includegraphics{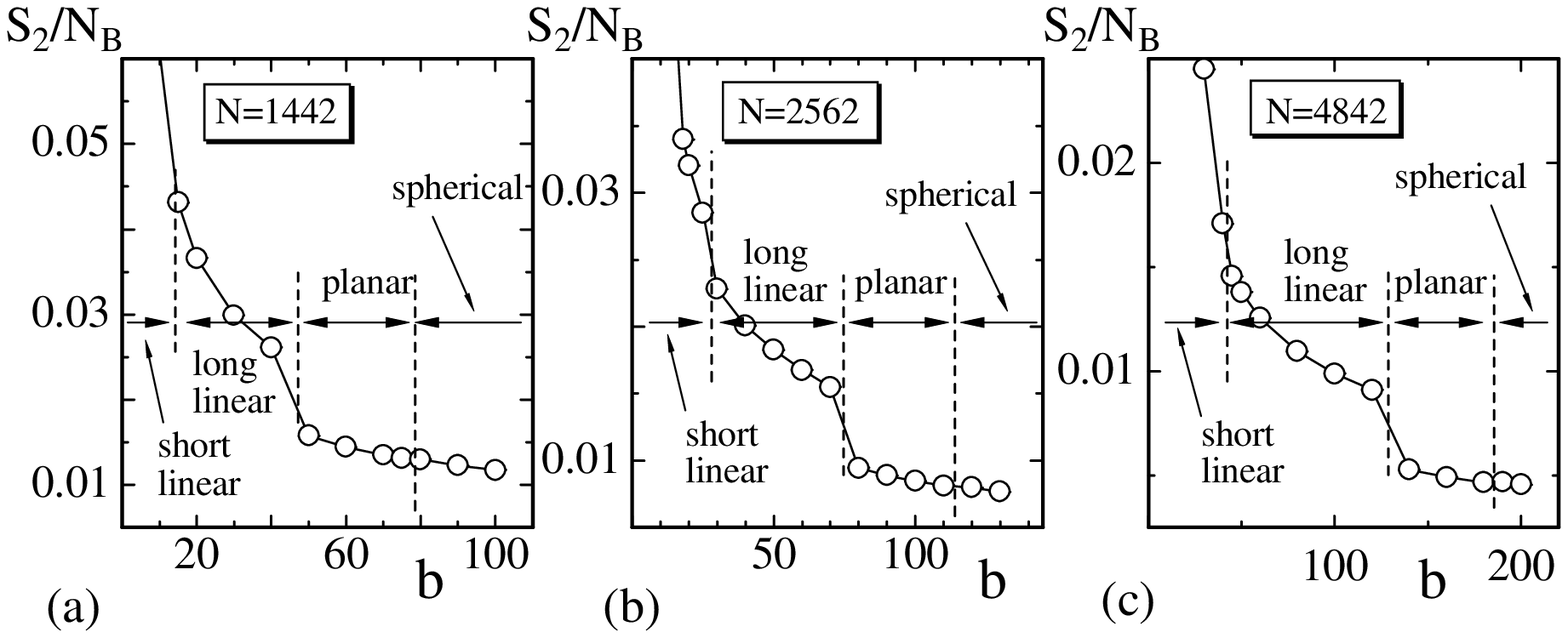}
}
\caption{The one-dimensional bending energy $S_2/N_B$ versus $b$ obtained on the surfaces of size (a) $N\!=\!1442$, (b) $N\!=\!2562$, and (c) $N\!=\!4842$. Dashed lines denote the phase boundaries.   }
\label{fig-7}
\end{figure*}
The one-dimensional bending energy $S_2/N_B$ versus $b$ is shown in Figs.\ref{fig-7}(a)--\ref{fig-7}(c), where the surface size is $N\!=\!1442$, $N\!=\!2562$, and $N\!=\!4842$.  The total number of combinations $1-{\bf t}_i \cdot {\bf t}_j$ in $S_2$ of Eq.(\ref{S1S2}) is identical to $N_B$ the total number of bonds. For this reason, $S_2$ is divided by $N_B$ in Figs.\ref{fig-7}(a)--\ref{fig-7}(c). The positions of the dashed lines are identical to those in the previous figures. The discontinuity is clear only at the boundary between the long linear phase and the planar phase.

 We also see that $S_2/N_B$ continuously varies at the boundary between the planar phase and the spherical phase. This is inconsistent to the naive expectation that physical quantities seem to have anomalous values at the edge, which is the singular part of the planar surface. The reason of the continuous behavior in $S_2/N_B$ is because the vertices are dense in the perimeter region. Triangles become very thin at the edge of the planar surface, then $1-{\bf t}_i \cdot {\bf t}_j$ remains quite small at the edge and is comparable to that in the central part of the planar surface or in the spherical surface. 
 
\begin{figure*}[hbt]
\centering
\resizebox{0.7\textwidth}{!}{%
\includegraphics{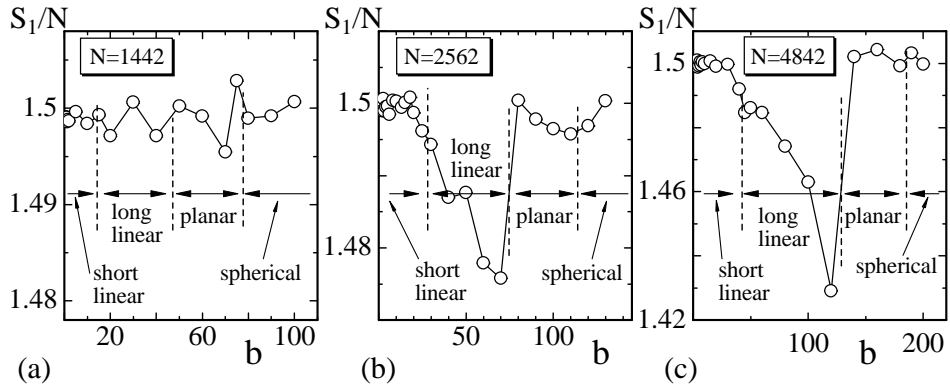}
}
\caption{The area energy $S_1/N$ versus $b$ obtained on the surfaces of size (a) $N\!=\!1442$, (b) $N\!=\!2562$, and (c) $N\!=\!4842$. Dashed lines denote the phase boundaries.    }
\label{fig-8}
\end{figure*}
Finally, we show the area energy $S_1/N$ in Figs.\ref{fig-8}(a)--\ref{fig-8}(c). It is expected that $S_1/N\!=\!3/2$ from the scale invariant property of the partition function. We find from Figs.\ref{fig-8}(a)--\ref{fig-8}(c) that the relation $S_1/N\!=\!3/2$ is satisfied except in the long and the short linear phases. The surface in the linear phases is almost one-dimensional and therefore, the random shift of the vertices $X$ in Monte Carlo updates seems not efficiently performed on such straight thin surfaces. For this reason, $S_1/N$ seems to deviate slightly from $3/2$ in the linear phases.

\section{Summary and conclusions}
We have studied shape transformation phenomena in a surface model of Nambu and Goto, where the mechanical strength is given by a one-dimensional bending energy. The lattice is a triangulated mesh and is identical to the one widely used for surface models with two-dimensional bending energy. The Nambu-Goto surface model is known as an ill-defined model even with the standard two-dimensional bending energy. Thus, we aimed at showing that the model turns to be well defined when the one-dimensional bending energy is included in the Hamiltonian. 

It was shown that the model is well defined; there appears no surface like a chestnut bur even at sufficiently small bending rigidity $b\to 0$. The surfaces have always a well-defined surface structure in the whole $b$ region including $b\!\to\!0$.

We have also shown that the model has a variety of shapes including the smooth spherical surface and the collapsed surface. There are 6 different phases in the model, and almost all two neighboring phases are separated by discontinuous transitions. The appearance of the planar phase should be emphasized, because it is not yet reported that the fixed-connectivity surface model has a planar phase. It should also be emphasized that surface fluctuations are hardly seen in the inflated phases such as the planar phase and the spherical phase. The surfaces in those phases appear to be quite smooth. 

The results reported in this paper are very different from those in \cite{KOIB-PLA-2007}, where the standard Gaussian bond potential is assumed as a bond potential. Bond potential considerably influences the phase structure of meshwork models.

\section*{Acknowledgment}
This work is supported in part by a Grant-in-Aid for Scientific Research from Japan Society for the Promotion of Science. 



\end{document}